\documentclass[11pt]{article}
\usepackage{hyperref}%
\pdfoutput=1%
\begin{document}%
\title{Oblique-shock/turbulent-boundary-layer interaction}%
\author{E. Touber$^1$\hspace{1.25cm}N.~D. Sandham$^2$\\%
\\$^1$Department of Aeronautics, Imperial College London,\\%
South Kensington Campus, London SW7 2AZ, U.K.\\~\\%
$^2$School of Engineering Sciences, University of Southampton,\\%
University Road, Southampton SO17 1BJ, U.K.}%
\date{}%
\maketitle%
\begin{abstract}%
The present numerical investigation uses well-resolved large-eddy simulations to study the low-frequency unsteady motions observed in shock-wave/turbulent-boundary-layer interactions. Details about the numerical aspects of the simulations and the subsequent data analysis can be found in three papers by the authors (Theo. Comput. Fluid Dyn., 23:79--107 (2009); Shock Waves, 19(6):469--478 (2009) and J. of Fluid Mech. (2011)). The fluid dynamics video illustrates the complexity of the interaction between a Mach 2.3 supersonic turbulent boundary layer and an oblique shock wave generated by a 8-degree wedge angle. The first part of the video highlights the propagation of disturbances along the reflected shock due to the direct perturbation of the shock foot by turbulence structures from the upstream boundary layer. The second part of the video describes the observed block-like back-and-forth motions of the reflected shock, focusing on timescales about two orders of magnitude longer than the ones shown in the first part of video. This gives a visual impression of the broadband and energetically-significant peak in the wall-pressure spectrum at low frequencies. The background blue-white colouring represents the temperature field (with white corresponding to hot) and one can clearly appreciate why such low-frequency shock motions can lead to reduced fatigue lifetimes and is detrimental to aeronautical applications.%
\end{abstract}%
\end{document}